\documentclass{iau}
\usepackage{graphicx}
\def\gtsima{$\; \buildrel > \over \sim \;$}
\def\ga{\lower.5ex\hbox{\gtsima}}	
\def\ltsima{$\; \buildrel < \over \sim \;$}
\def\la{\lower.5ex\hbox{\ltsima}}            

\title[IAUS 304.~~Deep radio surveys and radio emission in radio-quiet AGN] 
{The AGN content of deep radio surveys and radio emission in radio-quiet AGN.   
~~~~~~~Why every astronomer should care about deep radio fields}

\author[P. Padovani et al.]   
{P. Padovani$^1$, M. Bonzini$^1$, N. Miller$^2$, K. I. Kellermann$^{3}$, V. Mainieri$^{1}$, 
P. Rosati$^{1,4}$, P. Tozzi$^{5}$ \and S. Vattakunnel$^{6}$}

\affiliation{$^1$European Southern Observatory, Karl-Schwarzschild-Str. 2,
D-85748 Garching bei M\"unchen, Germany\\ email: {\tt ppadovan@eso.org} \\[\affilskip]
$^2$Department of Mathematics and Physical Sciences, Stevenson University, 
1525 Greenspring Valley Road, Stevenson, MD  21153-0641, USA\\[\affilskip]
$^{3}$National Radio Astronomy Observatory, 520 Edgemont Road, 
 Charlottesville, VA 22903-2475, USA\\[\affilskip]
$^{4}$ Dipartimento di Fisica e Scienze della Terra, Universit\`a di Ferrara, Via Saragat 1, 
I-44122, Ferrara, Italy (current address)\\[\affilskip] 
$^{5}$INAF - Osservatorio Astrofisico di Arcetri, Largo E. Fermi, I-50125, Firenze, Italy\\[\affilskip]
$^{6}$INAF, Osservatorio Astronomico di Trieste, Via G. B. Tiepolo 
11, I-34131, Trieste, Italy}

\pubyear{2014}
\volume{304}  
\pagerange{xxx--xxx}
\jname{Multiwavelength AGN Surveys and Studies}
\editors{A. Mickaelian, F. Aharonian \& D. Sanders, eds.}
\begin{document}

\maketitle

\begin{abstract}
We present our very recent results on the sub-mJy radio source populations at
1.4 GHz based on the Extended Chandra Deep Field South VLA survey, which
reaches $\sim 30~\mu$Jy, with details on their number counts, evolution, and
luminosity functions. The sub-mJy radio sky turns out to be a complex mix of
star-forming galaxies and radio-quiet AGN evolving at a similar, strong rate
and declining radio-loud AGN. While the well-known flattening of the radio
number counts below 1 mJy is mostly due to star-forming galaxies, these
sources and AGN make up an approximately equal fraction of the sub-mJy
sky. Our results shed also light on a fifty-year-old issue, namely radio
emission from radio-quiet AGN, and suggest that it is closely related to star
formation, at least at $z \sim 1.5 - 2$. The implications of our findings for
future, deeper radio surveys, including those with the Square Kilometre
Array, are also discussed.  One of the main messages, especially to non-radio
astronomers, is that radio surveys are reaching such faint limits that, while
previously they were mainly useful for radio quasars and radio galaxies, they are now detecting
mostly star-forming galaxies and radio-quiet AGN, i.e., the bulk of the
extragalactic sources studied in the infrared, optical, and X-ray bands.
\end{abstract}

\firstsection 
\section{Deep radio surveys}

\subsection{The modern radio sky}

The GHz radio bright ($\ga 1$ mJy) sky consists mainly of ``classical'' radio
sources, that is radio quasars and radio galaxies. These are active galactic
nuclei (AGN) whose radio emission is generated from the gravitational
energy associated with a supermassive black hole and emitted through
relativistic jets of particles as synchrotron radiation. Below 1 mJy there is
an increasing contribution to the radio source population from synchrotron
emission resulting from relativistic plasma ejected from supernovae
associated with massive star formation in galaxies. Star forming galaxies (SFG),
however, appear not to be the only component of the faint radio sky, at least down
to $\sim 50~\mu$Jy at a few GHz (e.g., \cite[Gruppioni et al. 2003]
{Gruppioni_etal2003}, \cite[Padovani et al. 2009]{Padovani_etal09},
\cite[2011]{Padovani_etal11}, \cite[Padovani 2011]{Padovani_etal2011},
\cite[Norris et al. 2013]{Norris_etal2013}), contrary to the (until recently)
most accepted paradigm. It turned out that there are still plenty
of AGN down there but of the more numerous, radio-fainter type, and therefore
not associated to radio quasars and radio galaxies. 
These AGN, in fact, are of the so-called radio-quiet (RQ) type.

Why should every astronomer care about all this? For a variety of
reasons: 1) radio emission, through the so-called ``radio-mode'' feedback,
appears to play a very important role in galaxy evolution (e.g., \cite[Croton
  et al. 2006]{Croton_etal06}); 2) RQ AGN reside typically in spiral
galaxies, which are still forming stars, and therefore are likely to provide
a vital contribution to our understanding of the AGN -- galaxy co-evolution
issue; 3) radio observations are unaffected by absorption and therefore are
sensitive to all types of AGN, independently of their orientation (i.e., type
1s and type 2s); 4) finally, and most importantly, the fact that by going
radio faint one starts to detect the bulk of the AGN population (and not only
the small minority of radio quasars and radio galaxies) means that radio
astronomy is no longer a ``niche'' activity but is extremely relevant to a
whole bunch of extragalactic studies.

\subsection{The issue of radio-quiet AGN}

A further point of general interest has to do with RQ AGN. Soon after the
discovery of quasars in 1963 it was realized that the majority of them were
not as strong radio sources as were the first quasars and were undetected
by the radio telescopes of the time: they were ``radio-quiet.'' It was later
realized that these sources were actually only ``radio-faint.'' For the same
optical power their radio powers were $\approx 3$ orders of magnitude smaller
than their radio-loud (RL) counterparts. RQ AGN were until
recently normally found in optically selected samples and are characterized
by relatively low radio-to-optical flux density ratios and radio powers. It
is important to realize that the distinction between the two types of AGN is
not simply a matter of semantics: the two classes represent intrinsically
different objects, with RL AGN emitting most of their energy over the entire 
electromagnetic spectrum non-thermally
and in association with powerful relativistic jets, while the multi-wavelength 
emission of RQ AGN is dominated by thermal emission, directly or indirectly 
related to the accretion disk.

The mechanism responsible for radio emission in RQ AGN has been a matter of
debate for the past fifty years. Alternatives have included a scaled down
version of the RL AGN mechanism (e.g., \cite[Miller et
  al. 1993]{Miller_etal93}, \cite[Ulvestad et al. 2005]{Ulvestad_etal05}),
star formation (\cite[Sopp \& Alexander 1991]{Sopp_91}), a black hole
rotating more slowly than in RL AGN (\cite[Wilson \& Colbert 1995]{Wilson_95}),
and many more. This is a non-trivial issue for various reasons: 1) most ($>
90\%$) AGN are RQ; 2) some of the proposed explanations have profound
implications on our understanding of AGN physics (jets, accretion, black hole
spin, etc.); 3) some others are very relevant for the relationship between
AGN and star formation in the Universe (related to ``AGN feedback''), which
is a hot topic in extragalactic research. We note that it is important to compare 
the properties of the two AGN classes in the band where they differ most, 
that is the radio band.

\section{The {\it Chandra} Deep Field South} 

This is exactly what we did. Our observations were done in the {\it Chandra} Deep Field 
South (CDFS) area, which is
part of the Great Observatories Origins Deep Survey (GOODS) and as such one
of the most intensively studied regions in the sky. The CDFS is a brainchild
of Riccardo Giacconi, who conceived the idea of having the {\it Chandra}
X-ray observatory stare at the same spot of the sky for a long time to reach
very faint X-ray fluxes. After the initial 1 Msec (11.6 days; \cite[Giacconi et
  al. 2002]{Giacconi_etal2002}), one more was added (\cite[Luo et
  al. 2008]{Luo_etal2008}). Four Msecs (1.5 months) were finally reached by \cite[Xue et
  al. (2011)]{Xue_etal2011}. More observing time has been granted in the
meantime so that a total of 7 Msecs (almost three months) of observing time
should be available by the end of 2014.

\cite[Kellermann et al. (2008)]{Kellermann_etal08} used the National Radio
Astronomy Observatory (NRAO) Very Large Array (VLA) to obtain 1.4 GHz data,
with 8.5~$\mu$Jy rms noise per 3.5" x 3.5" beam, in a field centered on the
CDFS, defining a complete sample of 198 radio sources sources reaching $\sim
43~\mu$Jy over 0.2 deg$^2$. These data were exploited in a series of papers
on optical counterparts (\cite[Mainieri et al. 2008]{Mainieri_etal08}), X-ray
properties (\cite[Tozzi et al. 2009]{Tozzi_etal09}), and source populations,
evolution, and luminosity functions (LFs) (\cite[Padovani et
  al. 2009]{Padovani_etal09}, \cite[2011]{Padovani_etal11}). \cite[Miller et
  al. (2008]{Miller_etal08}, \cite[2013)]{Miller_etal13} observed the
so-called Extended CDFS (E-CDFS), again using the VLA, reaching a somewhat
smaller rms over a larger area, namely $\sim 6~\mu$Jy rms noise, and
therefore $\sim 30~\mu$Jy at $5\sigma$, in a 2.8" x 1.6" beam over 0.3
deg$^2$. This resulted in a sample of almost 900 sources. Our group is
exploiting these new radio data with the aim of addressing the issues of 
the faint radio source population and RQ AGN in more detail,
given the larger and slightly deeper E-CDFS sample, as compared to the CDFS
one. In particular, \cite[Bonzini et al. (2012)]{Bonzini_etal2012} have
identified the optical and infrared (IR) counterparts of the E-CDFS sources,
finding reliable matches and redshifts for $\sim 95\%$ and $\sim 81\%$ of
them respectively, while \cite[Vattakunnel et
  al. (2012)]{Vattakunnel_etal2012} have identified the X-ray counterparts
and studied the radio -- X-ray correlation for SFG. 
Finally, \cite[Bonzini et al. (2013)]{Bonzini_etal2013} have provided
reliable source classification.

\subsection{The classification of faint radio sources}

The classification of faint radio sources is complex. First,
these objects are quite faint in the
optical/near IR regimes. The sources with an optical counterpart have a
median magnitude $R \sim 22.8$, but can be hosted in galaxies as faint as $R \sim 27$. 
Therefore, getting
spectra, which can be used for an optical classification is not feasible for
the bulk of the objects. But even if we had optical spectra for all our
sources we would still have problems since, as also remarked by a few
speakers at this conference (e.g., Laura Trouille) a single band only gives a
biased view of the properties of AGN and the optical band is the worst one,
being strongly affected by obscuration. Indeed, there are quite a few
examples of optically boring sources where the AGN is detected only in the
X-ray band (i.e., some of the so-called X-ray Bright Optically Normal
Galaxies; XBONGs). One then needs to use all the multi-wavelength information
available for the E-CDFS, which is substantial, to figure out what a radio
source really is, looking for AGN in the IR and X-ray bands as well.

\begin{figure}[ht]
\begin{center}
 \includegraphics[width=9cm]{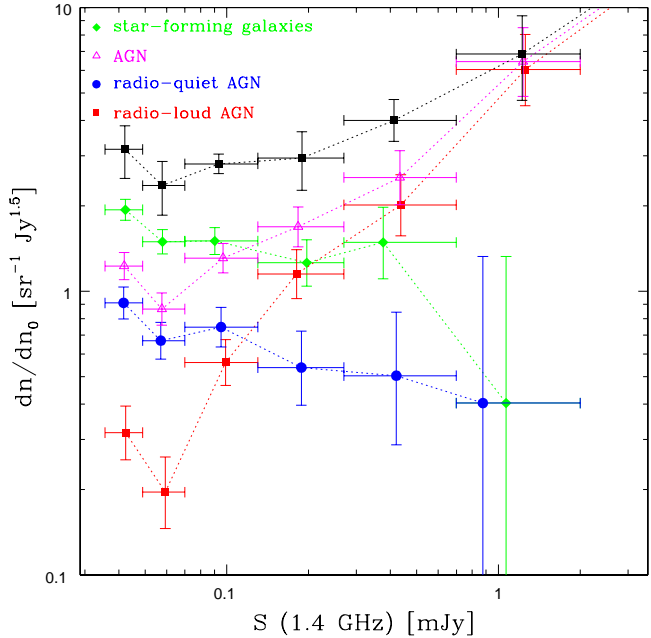} 
 \caption{Preliminary Euclidean normalized E-CDFS source counts: whole sample
   (black squares), SFG (green diamonds), all AGN (magenta triangles),
   radio-quiet AGN (blue circles), and radio-loud AGN (red squares).  Error
   bars correspond to $1\sigma$ Poisson errors (\cite[Gehrels
     1986]{Gehrels_86}).}
   \label{counts}
\end{center}
\end{figure}

\begin{figure}[ht]
\begin{center}
 \includegraphics[width=9cm]{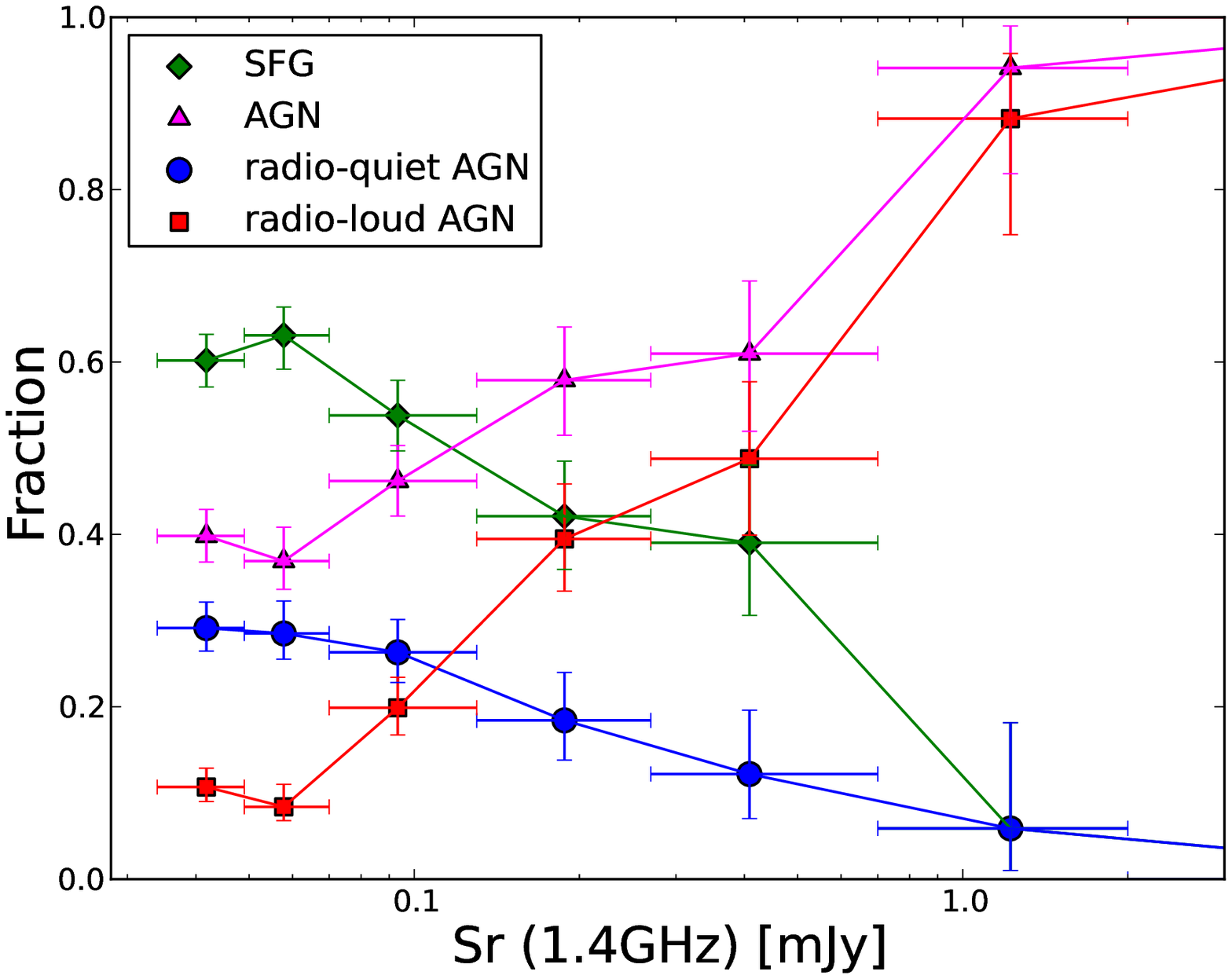} 
 \caption{Relative fraction of the various classes of E-CDFS sources as a function
 of radio flux density: SFG
   (green diamonds), all AGN (magenta triangles), radio-quiet AGN (blue
   circles) and radio-loud AGN (red squares). Error bars correspond to
   $1\sigma$ Poisson errors (\cite[Gehrels 1986]{Gehrels_86}). Adapted from
   \cite[Bonzini et al. (2013)]{Bonzini_etal2013}.}
   \label{fractions}
\end{center}
\end{figure}

\cite[Bonzini et al. (2013)]{Bonzini_etal2013} have used a modified version
of the classification scheme used by \cite[Padovani et al. (2011)]
{Padovani_etal11} to disentangle SFG, RQ, and RL AGN. In short, 
they identify RL AGN using the so-called $q_{24}$
parameter, which is defined by $\log f_{24 \mu m }/f_{1.4GHz}$, 
while RQ AGN are distinguished from SFG according to their IRAC (near-IR) 
colours and X-ray power. Finally, other
parameters were examined to check the classification and sort out
possible outliers. These included: the presence of an inverted radio spectrum
($\alpha_{\rm r} < 0$, where $S_{\nu} \propto \nu^{-\alpha}$), possible Very
Long Baseline Array (VLBA) detections, broad/high excitation lines in the
optical spectra, Polycyclic Aromatic Hydrocarbon (PAH) features, X-ray
absorption and variability.

\subsection{The faint radio source population} 

Thanks to this careful classification, we now have what we believe is the
most accurate determination of the source population of faint radio
sources. Fig. \ref{counts} shows the preliminary Euclidean normalized number
counts for our sample, along with the population subsets 
(Padovani et al., in preparation). 
Fig. \ref{fractions}, adapted from \cite[Bonzini et al. (2013)]
{Bonzini_etal2013}, plots the relative fractions of the various classes as a
function of radio flux density. One can see that AGN dominate at large flux
densities ($\ga 1$ mJy) but SFG become the main population below $\approx
0.1$ mJy. Similarly, RL AGN are the predominant type of AGN above 0.1 mJy but
their contribution drops fast at lower flux densities. In more detail, AGN make up
$43\pm4$\% of sub-mJy sources, going from 100\% of the total at $\sim 10$ mJy,  
down to 39\% at the survey limit. SFG, on the other hand, which represent 
$57\pm3$\% of the sub-mJy sky,
are missing at high flux densities but become the dominant population below
$\approx 0.1$ mJy, reaching 61\% at the survey limit. RQ AGN represent
$26\pm6$\% (or 60\% of all AGN) of sub-mJy sources but their fraction appears
to increase at lower flux densities, where they make up 73\% of all AGN and
$\approx 30$\% of all sources at the survey limit, up from $\approx 6$\% at
$\approx 1$ mJy. These results are in good agreement with those of
\cite[Padovani et al. (2011)]{Padovani_etal11}. The strong message is that
below 0.1 mJy the radio sky, which at large flux densities is characterized
by the prevalence of synchrotron radiation associated with powerful jets,
is instead dominated by star-formation-related processes.

\subsection{Evolution and luminosity functions}  

Having redshifts for the majority of our sources, we can also study the
LFs and evolution of faint radio sources. The most general approach to do
this is to perform a maximum likelihood fit of an evolving LF 
to the observed distribution in luminosity and redshift (see details
in \cite[Padovani et al. 2011] {Padovani_etal11}). This method makes maximal
use of the data and is free from arbitrary binning, although it is model
dependent.

\begin{figure}[ht]
\begin{center}
 \includegraphics[width=8cm]{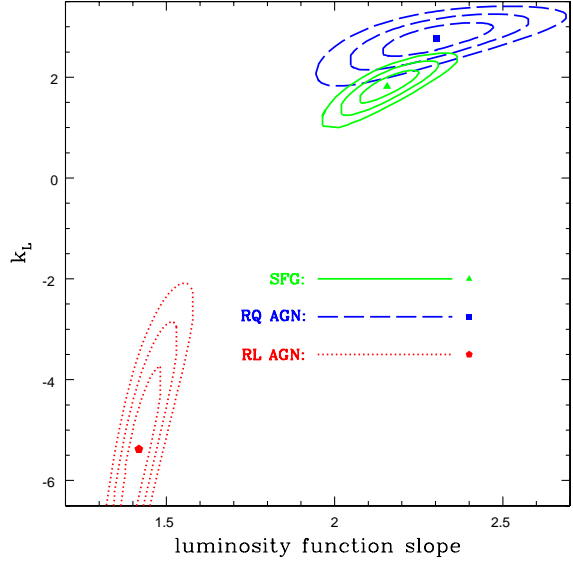} 
 \caption{Preliminary maximum likelihood confidence contours ($1\sigma$,
   $2\sigma$, and $3\sigma$) for the E-CDFS sample assuming a pure luminosity 
   evolution of the type
   $\propto (1+z)^{k_L}$ and a single power-law for the LF. 
   The best fit values for the various classes are indicated by
   the different symbols.}
   \label{contours}
\end{center}
\end{figure}

Fig. \ref{contours} plots preliminary maximum likelihood confidence contours
($1\sigma$, $2\sigma$, and $3\sigma$) under the simple assumptions of 
a pure luminosity evolution of
the type $\propto (1+z)^{k_L}$ and a single power-law for the LF. 
$k_L > 0$ indicates positive evolution, i.e., the higher the redshift the 
larger the power, while $k_L < 0$ indicates negative evolution, i.e., the higher 
the redshift the smaller the power. RQ AGN and SFG occupy the same region of parameter space (strong
positive evolution and steep LF), while RL and RQ AGN are totally disconnected, the
former displaying negative evolution and a flat LF. Note that, even considering a more 
complicated picture, as suggested by a more detailed analysis of the data (i.e., a dual 
power-law LF: Padovani et al. in prep.), RQ 
AGN and SFG still evolve similarly in the radio band. In particular, the
derivation of the local LF confirms that RL AGN have a much flatter LF than
RQ ones and shows that the RQ AGN LF seems to be an extension of the SFG LF
at higher radio powers (see also \cite[Padovani et al. 2009]
{Padovani_etal09}).  

\section{Radio emission in radio-quiet AGN}

Our results suggest very close ties between star formation and radio emission
in RQ AGN at $z \sim 1.5 - 2$, the mean redshift of our sources, since their
evolution is similar to that of SFG and their LF appears to be an extension
of the SFG LF. Furthermore, radio emission in the two classes of AGN appears 
to have a different origin. If RQ AGN were simply ``mini RL'' AGN, in fact,
they would have to share the evolutionary properties of the latter and their
LF should also be on the extrapolation of the RL LF at low
powers. Neither of these is borne out by our data. This conclusion
agrees with our previous results (\cite[Padovani et al. 2011]
{Padovani_etal11}) and \cite[Kimball et
  al. (2011)]{Kimball_etal2011}. This result is further confirmed by the 
comparison of the star formation rates (SFRs) derived from the far-IR 
and radio luminosities, assuming that all the radio emission is due to SF. 
For RQ AGN and SFG the two SFR estimates are fully consistent, while 
for RL AGN the agreement is poor due to the large contribution of the
relativistic jet to their radio luminosity (Bonzini et al., this volume).

\section{A revolution in radio astronomy}

The Square Kilometre Array (SKA; http://www.skatelescope.org) 
will offer an observing opportunity extending well into
the {\it nanoJy} regime with unprecedented versatility (we stress that at
present we have reached $\sim 15~\mu$Jy at 1.4 GHz). First science with the
so-called SKA$_1$ is scheduled early in the next decade. The
LOw Frequency ARray (LOFAR) has already started operations and will carry out large
area surveys at 15, 30, 60, 120 and 200 MHz (\cite[Morganti et
  al. 2009]{Morganti_etal09}), opening up a whole new region of parameter
space at low radio frequencies. The VLA has been upgraded and the new
instrument, the Jansky Very Large Array (JVLA; https://science.nrao.edu/facilities/vla), 
has hugely improved
capabilities and MERLIN, which is now e-MERLIN (http://www.e-merlin.ac.uk/), 
has greatly improved sensitivity. 
Many other radio
telescopes are currently under construction in the lead-up to the SKA
including the Australian Square Kilometre Array Pathfinder (ASKAP;
http://www.atnf.csiro.au/projects/askap/), Apertif
(http://www.astron.nl/\-general/\-apertif/apertif), and MeerKAT
(http://www.ska.ac.za/\-meerkat). These projects will survey the sky 
faster than has been possible with existing radio telescopes producing surveys
covering large areas of the sky down to fainter flux densities than presently
available (see \cite[Norris et al. 2013]{Norris_etal2013} for details).

As an example, simple modelling based on our results shows that a radio survey reaching $\sim
1~\mu$Jy at 1.4 GHz will need to cover only $\sim 30$ deg$^2$ to detect a
number of ``potential" AGN larger than those currently known, while $\sim
200$ deg$^2$ should reveal about one million such sources. The Evolutionary
Map of the Universe (EMU), one of the ASKAP projects, expects to reach flux
densities similar to those of the E-CDFS over $\sim 3/4$ of the sky,
detecting more than 70 million sources, about half of which will be
``potential" AGN (\cite[Norris et al. 2011]{Norris_etal2011}). The
``potential'' here is important because, as detailed above, the
classification of faint radio sources requires a great deal of ancillary,
multi-wavelength information, which will not be easy to get at very faint
levels (\cite[Padovani 2011]{Padovani_11}) or over very large areas.

\section{Conclusions}

We have shown that the GHz sub-mJy sky is a complex mix of evolving
star-forming galaxies and radio-quiet AGN and negatively evolving radio-loud
AGN. Contrary to what used to be the predominant view, AGN still make up
$\sim 40\%$ of the faint radio sky, although the deep ($< 0.1$ mJy) radio sky
appears to be dominated by star forming galaxies and therefore stellar
processes. Radio-quiet AGN are also very important, accounting for $\sim
60\%$ of all sub-mJy AGN and $\sim 25\%$ of the total radio faint population.
Radio surveys, which were in the past only useful to study radio-loud AGN,
have reached so deep that they are now dominated by the same galaxies detected
by IR, optical and X-ray surveys. As a result, the next generation radio
surveys, which will produce samples of tens of millions of sources, will be
an increasingly important component of multiwavelength studies of galaxy
evolution. Finally, by studying the evolution and luminosity functions of
faint radio sources, we have found a close relationship between star
formation and radio emission in radio-quiet AGN at $z \sim 1.5 - 2$, which is
confirmed by the very similar star formation rates obtained independently in
the radio and far-IR bands.



\end{document}